\documentclass[12pt]{iopart}
\usepackage{epsf}

\begin{document}

\title[]{Energy and System Size Dependence of Strangeness Production from SPS to RHIC}

\author{J. Takahashi for the STAR Collaboration}
\address{Instituto de F\'{i}sica Gleb Wataghin, Universidade Estadual de Campinas, Brazil}
\ead{jun@ifi.unicamp.br}

\begin{abstract}
Strange particle production is an important experimental observable that allows the study 
of the strongly interacting matter created in relativistic heavy-ion collisions. 
The STAR experiment at RHIC has a unique capability of measuring identified 
strange particles over a wide range of acceptance providing a rich set of data to
perform a systematic study. In addition to the data from Au+Au collisions, strange particles 
from p+p and d+Au collisions are also available for comparison and normalization. A new set of 
data from Cu+Cu reactions at 62 GeV and 200 GeV provides the chance to compare the system 
size dependence observed in Au+Au collisions with this smaller system size. In addition to the 
comparison of the yields, a statistical thermal model was used to extract freeze-out characteristics
for the different system sizes and collision energies. 
\end{abstract}

% \maketitle

\section{Introduction}

Relativistic heavy-ion collisions provide the ideal environment with which to study strongly interacting nuclear 
matter at different thermodynamic conditions. Lattice QCD calculations predict that above a certain critical 
temperature, nuclear matter undergoes a phase transition forming a deconfined state of quarks and  gluons~\cite{QCD}.
Various signatures of QGP formation have been proposed with the enhancement of strangeness production being
one of the experimental observables~\cite{Rafa01,Rafa02}. Despite the large amount of different signatures 
that were proposed, due to the complexity of the different processes that might be competing in the reaction, 
a clear and unique signal of the QGP formation is not possible. 
Thus, to understand and characterize the system formed in these heavy-ion collisions, it is necessary to perform 
a systematic study of the production of various particles including energy dependence and system size dependence 
as well as comparison to other reactions such as p+p, d+Au and Cu+Cu.  This kind of complete study is only possible 
now due to the large amount of data accumulated from several years of RHIC-BNL operations and the new analysis 
from the lower energy CERN-SPS experiments. 

STAR~\cite{star0} has a complex system of detectors specialized in measuring the maximum number of particles produced 
in a RHIC collision on an event by event basis. In particular, the tracking detectors from STAR have a high
efficiency for measuring charged particle tracks at mid-rapidity. The excellent track resolution and 
primary vertex position resolution yields a high efficiency for reconstructing strange particles through their
secondary vertex and cascade decay topologies.
In general, the invariant mass peaks of these particles can be extracted with good signal to noise ratios by
applying only some basic geometrical and track quality cuts. Energy loss (dE/dx) information of the daughter tracks is also
used to reduce combinatorial background. The background under the invariant mass 
peak can be subtracted by fitting the background shape or by simply counting the neighboring bins. 
Transverse momentum spectra for these strange particles are obtained by calculating the invariant mass for 
each $p_{T}$ bin and correcting for inefficiency and acceptance on a point by point basis. The efficiency and
acceptance corrections are calculated using simulated particles embedded into real events. The total yield (dN/dy)
is obtained by using Maxwell-Boltzmann or exponential curves to extrapolate into the low and high $p_{T}$ region
that is not covered by the measurement. Systematic errors are estimated considering several factors in the different
steps of the analysis such as the variation in the yield due to variations in the background subtraction techniques, 
variations due to differences in the topological geometric cuts and differences between real data and simulated 
data in the efficiency calculation.

Here in this article, we present an analysis that concentrates on the production of strange particles,
with special interest in the energy and system size dependence comparing the particle yields with predictions
obtained from a statistical thermal model. Data presented here are a compilation of STAR experiment data from the last 
seven years~\cite{star1,star2,Jun} and some additional data points in the lower energy range from experiments 
NA49~\cite{NA49} and NA57~\cite{NA57} obtained from the literature.

\section{Size Dependence}

\begin{figure}
\centerline{\epsfxsize 2in \epsffile{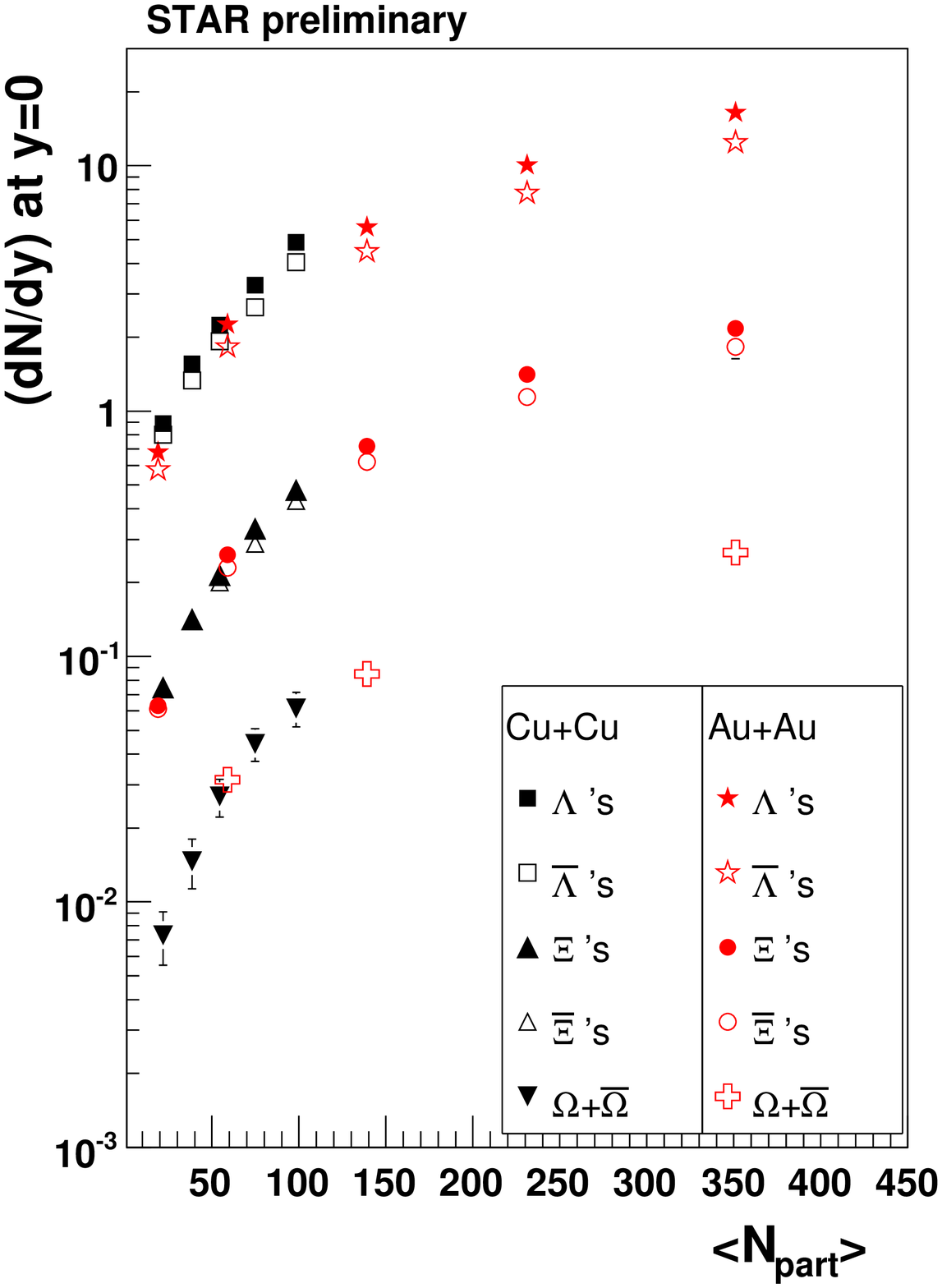} \epsfxsize 2in \epsffile{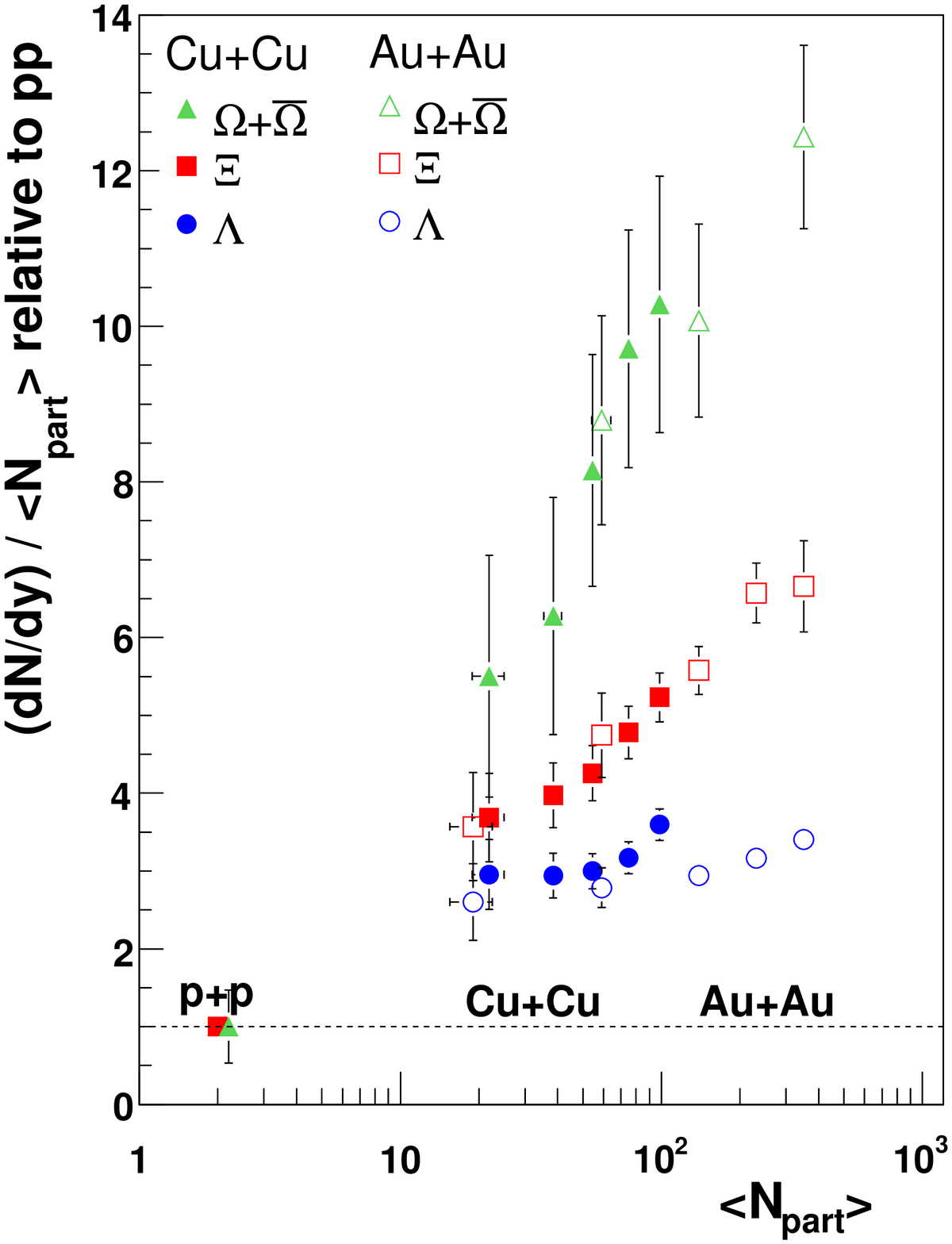} \epsfxsize 2in \epsffile{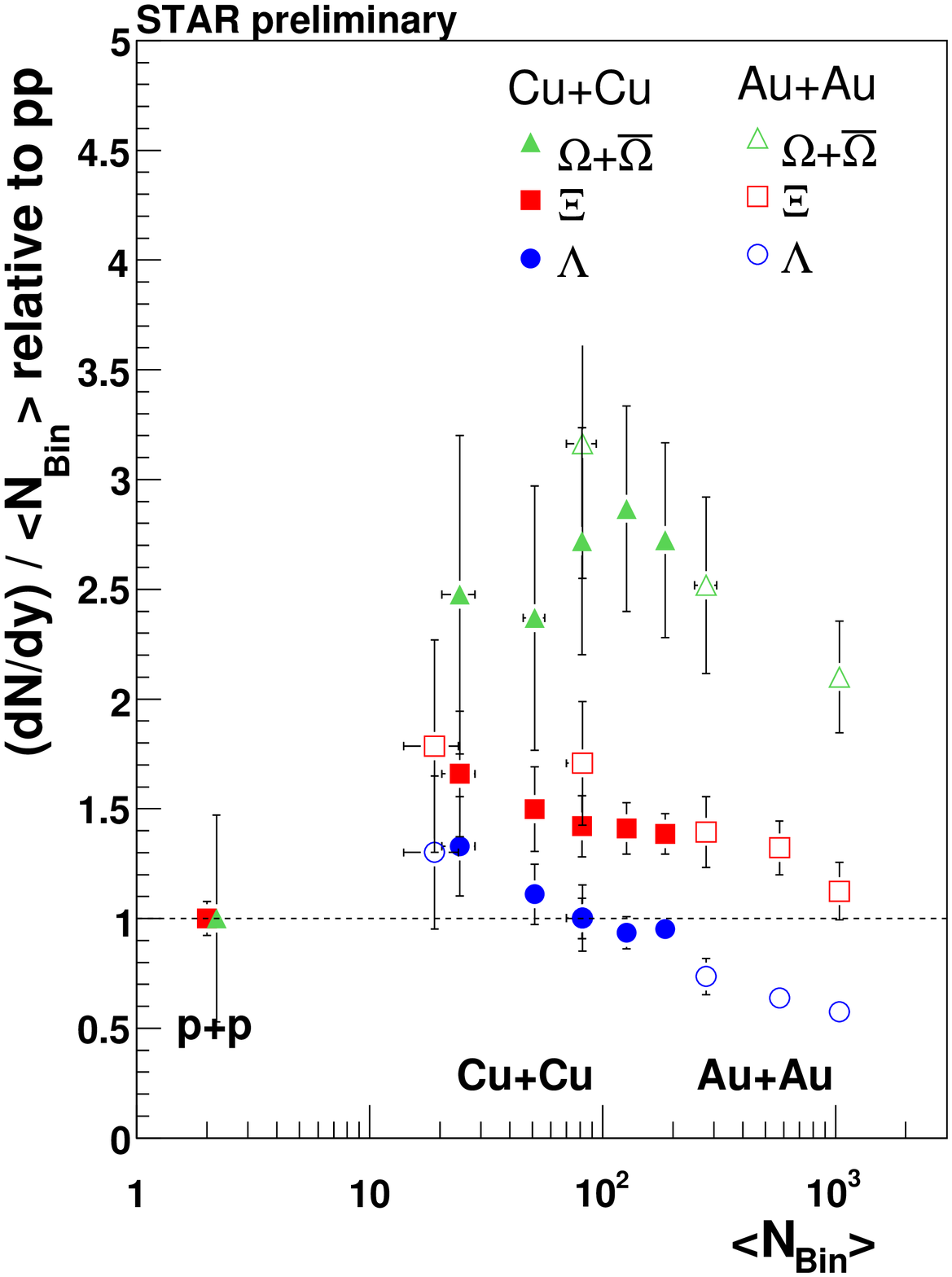} }
\caption{\label{label1} a) Strange particle yields for $\Lambda$, $\bar{\Lambda}$, $\Xi^{-+}$ and $\Omega^{-+}$ as 
a function of $\langle N_{part}\rangle$ for Au+Au and Cu+Cu collisions. Error bars are statistical only. 
b) Yields normalized by the collision number 
of participants $\langle N_{part}\rangle$ and c) number of binary collisions $\langle N_{bin}\rangle$.}
\end{figure}

The large amount of data measured for each reaction system also allows for the classification of events based on
the centrality of the collision. Collision centrality is determined by the measured charged particle multiplicities.
For each centrality one can estimate the average number of particles that participate in the reaction $\langle N_{part}\rangle$ 
using a Glauber Model calculation, that also provides the equivalent number of binary collisions $\langle N_{bin}\rangle$.
Figure~\ref{label1}a shows the yields of $\Lambda$, $\bar{\Lambda}$, $\Xi^{\pm}$ and $\Omega^{\pm}$ as a function 
of $\langle N_{part}\rangle$ for all the studied centralities in Au+Au and Cu+Cu collisions at $\sqrt{s_{NN}}=200$ GeV. 
A detailed discussion on $K^{0}_{S}$ and $\Lambda$ production in Au+Au and Cu+Cu collisions as a function of 
system size can be found elsewhere~\cite{Jun,Ant}. Most of the  Cu+Cu data seem to be consistent with Au+Au yields when 
compared using the system volume ($\langle N_{part}\rangle$). The yields of $\Lambda$ hyperons for the most central Cu+Cu data
seem to be slightly higher than the equivalent Au+Au data (around $30\%$). A possible explanation for this 
behavior could be found with a different scaling. A more detailed discussion was presented elsewhere at this conference~\cite{Ant}.

To evaluate the strangeness enhancement in heavy-ion collisions, Au+Au data is compared to p+p collisions 
normalized by $\langle N_{part}\rangle$. Figure~\ref{label1}b shows the normalized yields as a function of $\langle N_{part}\rangle$ for 
all centralities in Au+Au and Cu+Cu collisions.
A clear enhancement of hyperon production in Au+Au and Cu+Cu compared to p+p can be seen. Anti-Lambda and Anti-Xi yields are
not shown in this comparison but similar curves are observed, with slightly lower enhancement levels, which is 
expected due to non-zero net baryon number. Omega and Anti-Omega yields were added to reduce statistical error.
It is interesting to note that this enhancement can be observed 
even in the most peripheral bins in Au+Au and Cu+Cu compared to the p+p yields. In addition, the enhancement 
appears to increase linearly as a function of $\langle N_{part}\rangle$ and also with strange quark content. The strangeness 
enhancement hierarchy is in accordance to what is expected from Canonical thermal production~\cite{Redlich}, 
but the dependence with volume ($\langle N_{part}\rangle$) disagrees with this same picture of Canonical thermal production.

Figure~\ref{label1}c shows the strange particle yields normalized by the number of binary collisions $\langle N_{bin}\rangle$ and
compared to p+p. Once more, Cu+Cu data seems to be in agreement with the Au+Au data but with a smaller discrepancy 
for the $\Lambda$ yield in the most central Cu+Cu bin. This plot also shows that strange particle production 
does not scale with $\langle N_{bin}\rangle$, suggesting that a different scaling is required. 

\begin{figure}
\centerline{\epsfxsize 2.in \epsffile{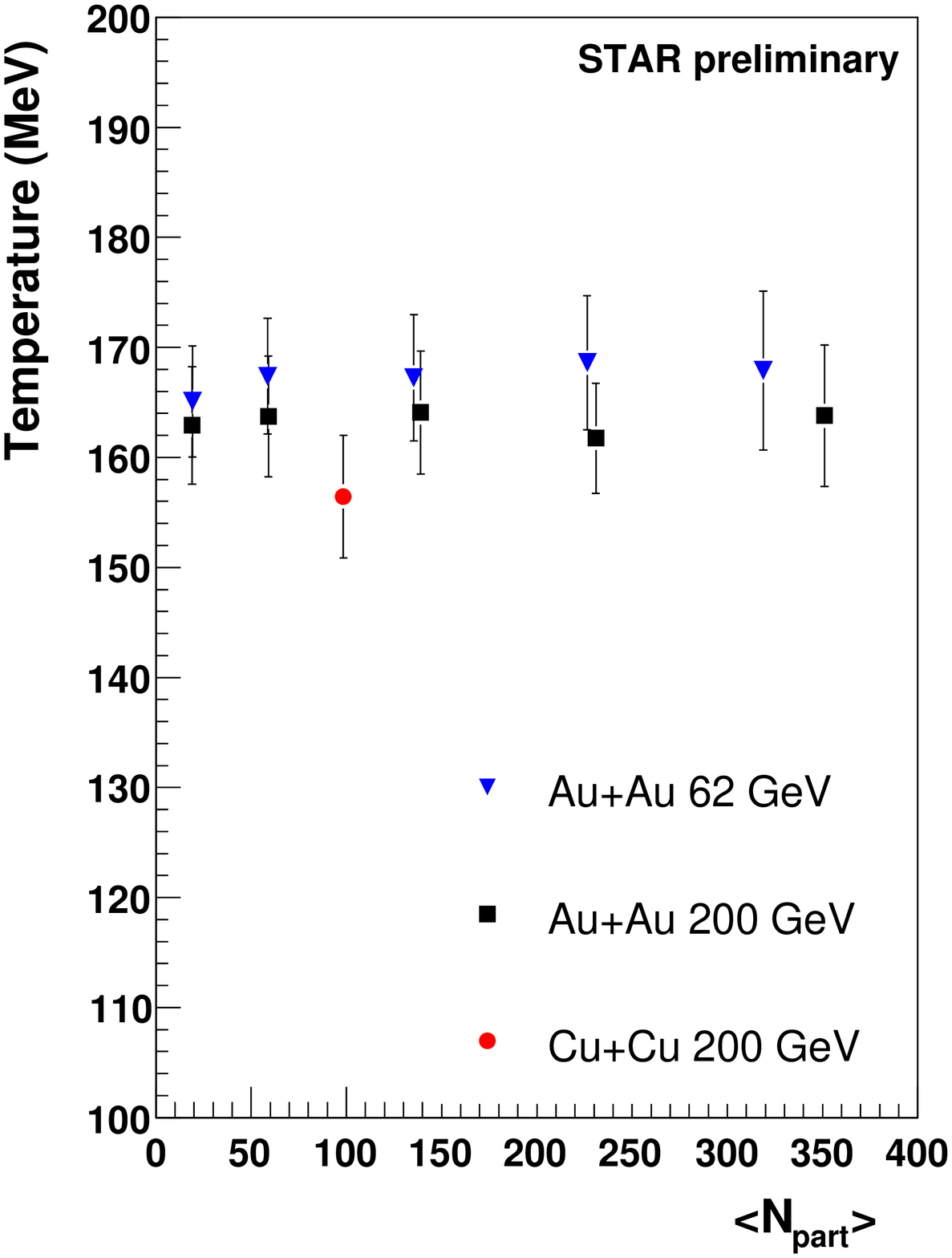} \epsfxsize 2.in \epsffile{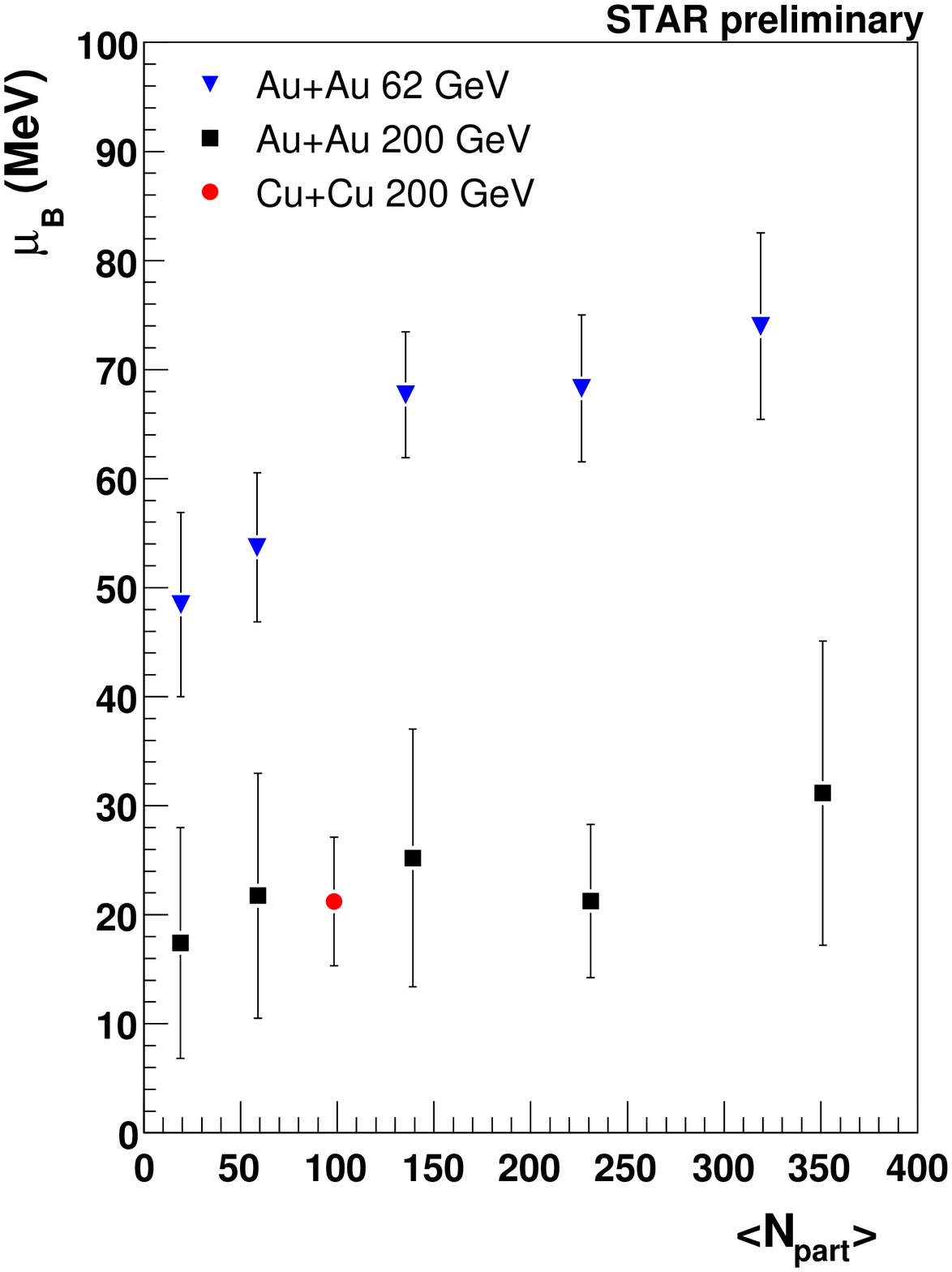} \epsfxsize 2.in \epsffile{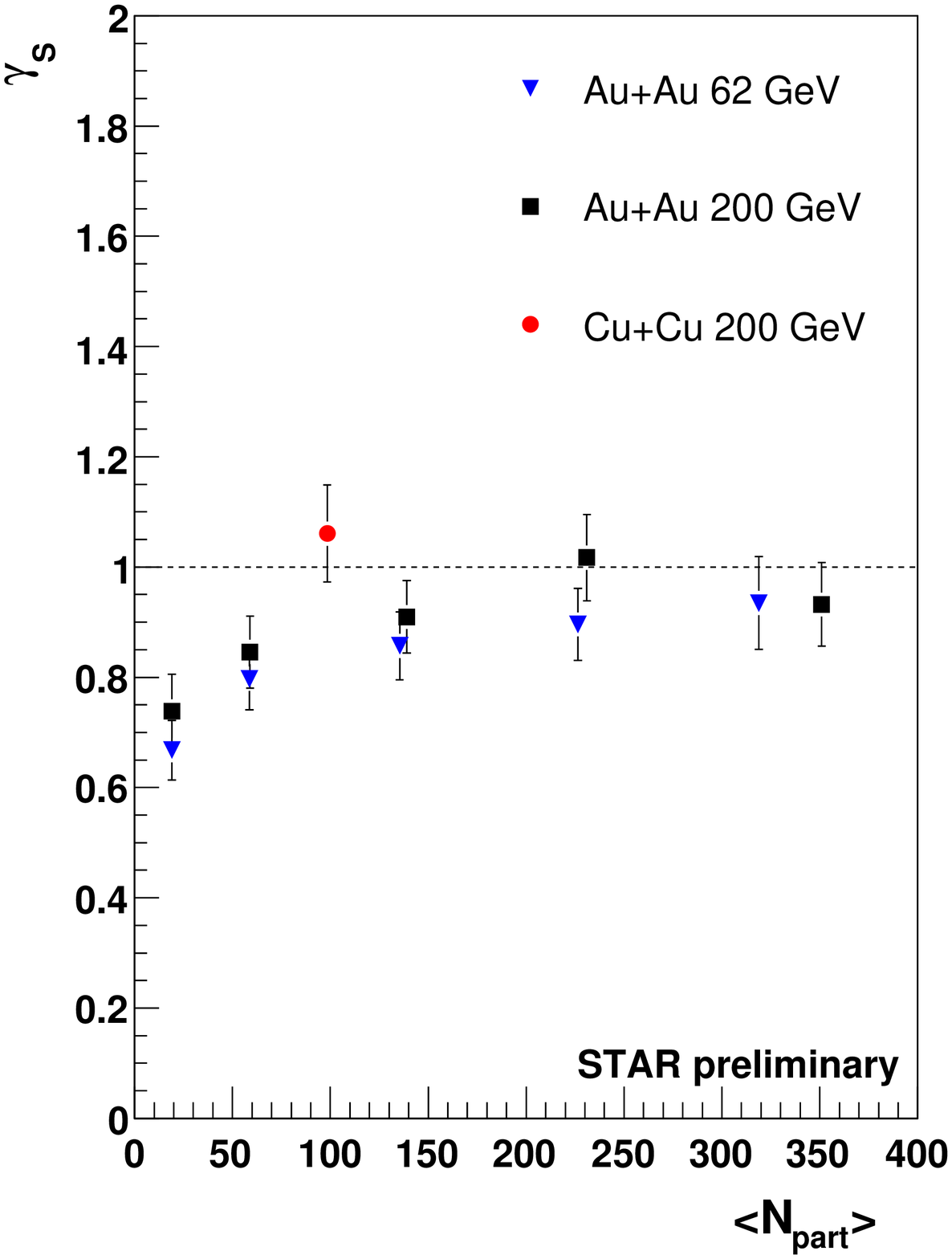} }
\caption{\label{label2} Results from a statistical thermal model fit \cite{Cleymanns}  
for Au+Au 62.4 GeV and 200 GeV and Cu+Cu 200 GeV (most central bin only). 
Chemical freeze-out temperature, baryon chemical potential and 
strangeness saturation parameter are shown as a function of $\langle N_{part}\rangle$.}
\end{figure}

A statistical thermal model \cite{Cleymanns} was used to fit the different particle ratios measured in Au+Au
and Cu+Cu collisions at energies of 200 GeV and 62.4 GeV, for different centrality bins. Yields of protons, pions
kaons, $\Lambda$'s and $\Xi$'s were used to calculate the yields for the fit.
Figure~\ref{label2} shows the results from these fits with chemical freeze-out temperature ($T_{ch}$), 
Baryon chemical potential ($\mu_{B}$) and strangeness saturation parameter ($\gamma_{S}$). The temperature
seems to show no dependence on the system size nor on the energy. The baryon chemical potential seems to be constant 
with sytem size for the 200 GeV Au+Au data, but shows a small increase with centrality in the lower energy 62.4 GeV Au+Au data. 
The strangeness saturation parameter, in the data from both energies, shows an increase with the system size till it reaches 
a saturation point.
The new Cu+Cu data, was analyzed only for the most central bin, and the comparison in the last plot indicates that
strangeness is already saturated for this system, which is not true in the equivalent $\langle N_{part}\rangle$ data from Au+Au 
collisions.

\section{Energy Dependence} 

\begin{figure}
\centerline{\epsfxsize 2.5in \epsffile{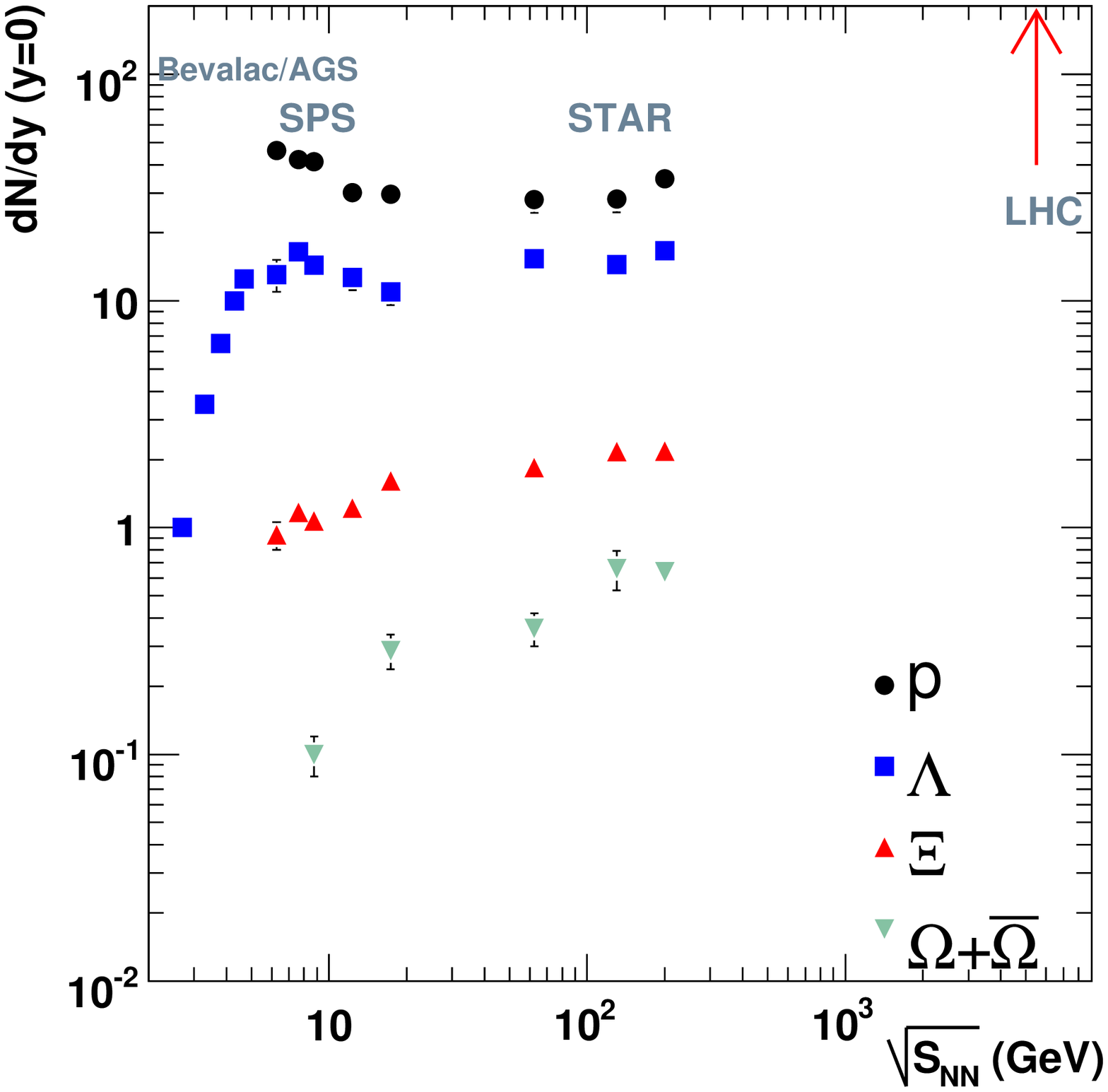} \epsfxsize 2.5in \epsffile{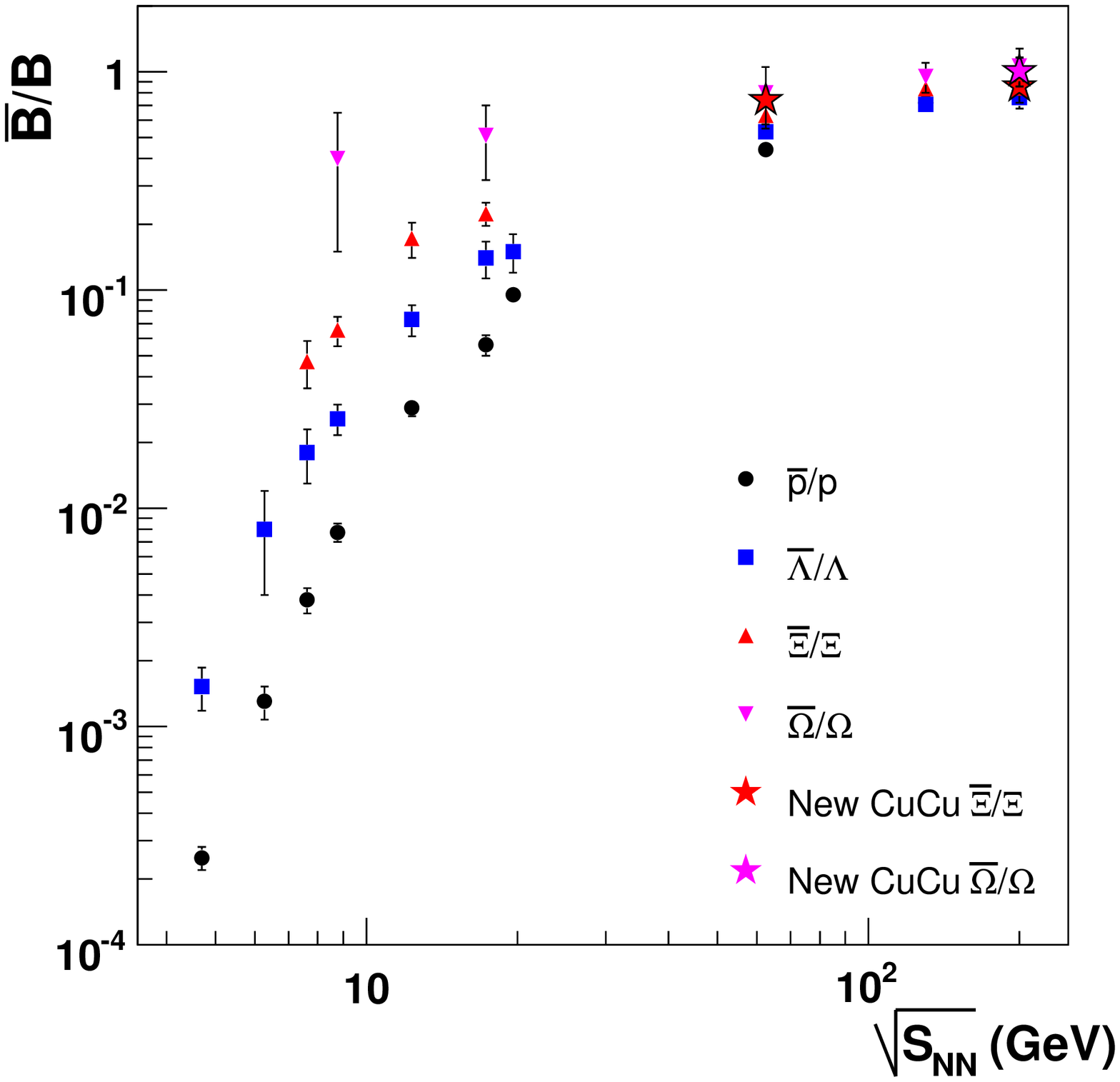} }
\caption{\label{label3} a) Baryon yields as a function of collision energy compiling data from AGS, SPS and RHIC 
experiments. b) Anti-baryon over Baryon ratio as a function of collision energy. New $\bar{\Xi}/\Xi$ and $\bar{\Omega}/\Omega$ 
ratios measured in Cu+Cu collisions at 62.4 GeV and 200 GeV are included in this systematic and shown as star symbols.}
\end{figure}

Figure~\ref{label3}a shows the particle yields as a function of collision energy compiling data from experiments at 
AGS-BNL~\cite{AGS}, SPS-CERN~\cite{NA49} and RHIC-BNL~\cite{PLB567,PRL86}. Baryon yields show a strong increase in the region of the AGS 
collision energies up to around 10 GeV, after which, the baryon yields stay relatively constant. 
Above SPS energies, there seems to be a smooth increase of the baryon production at higher energies. 
The anti-baryon yields increase continuously and smoothly with energy, that can be inferred from the ratio plots 
shown in figure~\ref{label3}b. These plots show the variation of the baryon production mode from a transport dominated 
regime into a pair-production regime. The new Cu+Cu data is included in the ratio plot with a star symbol for 62.4 GeV 
and 200 GeV, and show no difference to the Au+Au ratios.
From these curves, it is straightforward to predict the expected yields of these particles at the LHC energies.
For $\Lambda$ the expected yield is between 10 and 30, for $\Xi$ it is between 3 and 6, while for $\Omega$ 
between 0.4 and 0.7.

\begin{figure}
\centerline{\epsfxsize 2.5in \epsffile{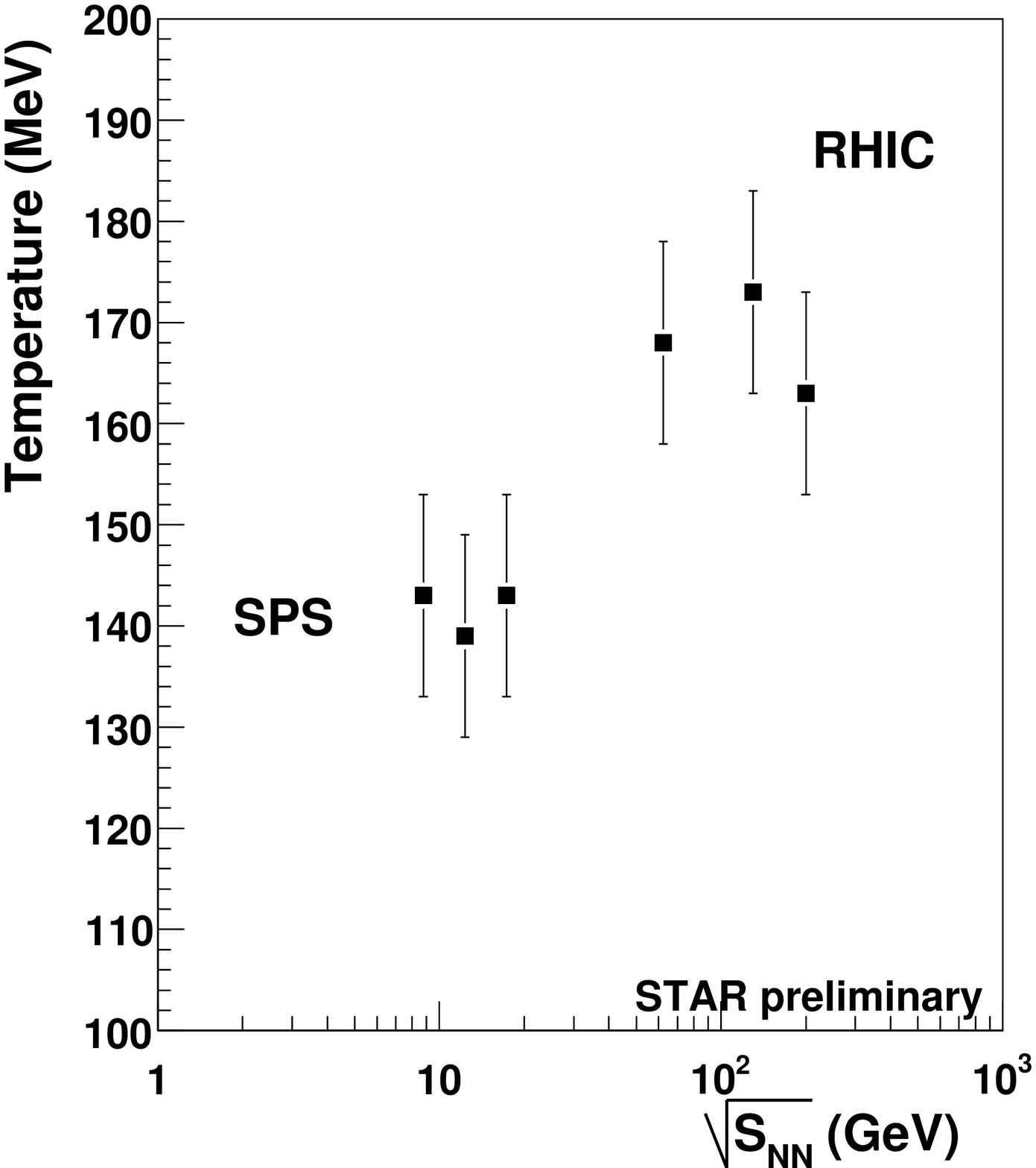} \epsfxsize 2.5in \epsffile{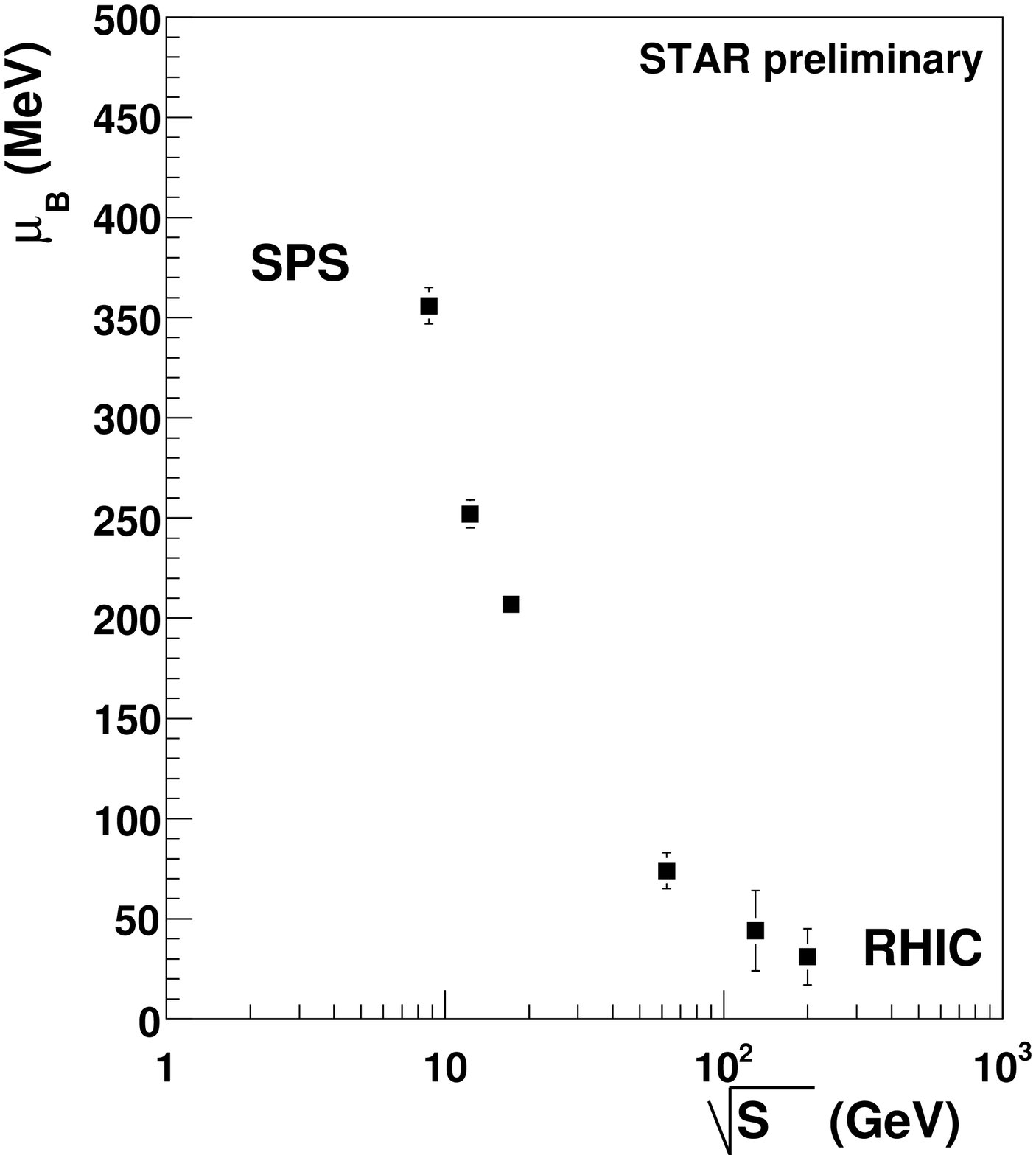}}
\caption{\label{label4} Thermal model fit parameters as a function of collision energy, for SPS and RHIC data points.}
\end{figure}

Figure~\ref{label4} shows an excitation function of the thermal fit parameters ($T_{ch}$,$\mu_{B}$) for the data 
from the NA49 experiment at the SPS-CERN and STAR-RHIC data. At RHIC energies, the freeze-out temperatures seem to have achieved a
saturation point around 170 MeV. All fits were done using the THERMUS thermal code ~\cite{Cleymanns}, using particle ratios calculated
from mid-rapidity yields (dN/dy). Once more, from
these plots, it is possible to predict the conditions expected from the future LHC heavy-ion program, where $\mu_{B}$ should
be close to zero and the temperature still around 170 MeV.

\section{Conclusions} 
There are now considerable amounts of data from heavy-ion reactions that allow for a systematic study of the
strange particle production for different energies and different reaction system sizes. This kind of study,
and in particular that of strange particle production, is crucial to understand the characteristics of the strongly 
interacting matter formed in RHIC collisions. The comparison between the new Cu+Cu data with the Au+Au data was shown 
and seems to be consistent with an $N_{part}$ scaling, except for the most central bin of the Lambdas. Thus it becomes 
clear that it is important to find the correct way to compare and scale the various different measurements from different 
reaction species and energies. 
The sharp change in behaviour of the particle yields excitation function in the lower energy region has been proposed 
as a signature for QGP formation~\cite{Horn}.  This shows the importance of a high-statistics energy scan as proposed 
in the next RHIC operation phase. Also, the extrapolation of the various measured curves to the higher energy range 
shows the importance of the new measurements expected from LHC.

\section*{Acknowledments} 
We wish to thank Funda\c{c}\~ao de Amparo a Pesquisa do Estado de S\~ao Paulo, and Conselho Nacional de 
Pesquisa de Desenvolvimento, Brazil for the support to participate in the SQM2007 conference.
We thank the RHIC Operations Group and RCF at BNL, and the
NERSC Center at LBNL and the resources provided by the
Open Science Grid consortium for their support. This work was supported
in part by the Offices of NP and HEP within the U.S. DOE Office 
of Science; the U.S. NSF; the BMBF of Germany; CNRS/IN2P3, RA, RPL, and
EMN of France; EPSRC of the United Kingdom; 
the Russian Ministry of Sci. and Tech.; the Ministry of
Education and the NNSFC of China; IRP and GA of the Czech Republic,
FOM of the Netherlands, DAE, DST, and CSIR of the Government
of India; Swiss NSF; the Polish State Committee for Scientific 
Research; Slovak Research and Development Agency, and the 
Korea Sci. \& Eng. Foundation.

\end{document}